\documentclass[aip,reprint]{revtex4-1}

\usepackage{fdsymbol}
\usepackage{setspace}
\usepackage[utf8]{inputenc}
\usepackage[T1]{fontenc} 

\usepackage{xcolor,color,soul} 
\usepackage{graphicx}
\usepackage{bm}
\usepackage{color}
\usepackage{longtable}
\usepackage{wrapfig}
\usepackage{natbib}
\usepackage[hidelinks, breaklinks=true, colorlinks=true, citecolor=blue, urlcolor=blue, linkcolor=blue]{hyperref}
\usepackage{framed}
\colorlet{shadecolor}{yellow}

\draft 

\begin{document}


\title{The elusive fluid-and-crystal coexistence state in simulations of monodisperse, hard-sphere colloids} 



\author{J. Galen Wang} 
\author{Umesh Dhumal} 
\affiliation{Mechanical and Aerospace Engineering, University of Missouri, Columbia, MO 65211, USA}
\author{Monica E. A. Zakhari} 
\affiliation{Department of Mechanical Engineering, Eindhoven University of Technology, Gemini-Zuid, Eindhoven, 5600 MB, The Netherlands}
\author{Roseanna N. Zia} 
\email{rzia@missouri.edu}
\affiliation{Mechanical and Aerospace Engineering, University of Missouri, Columbia, MO 65211, USA}


\date{\today}

\begin{abstract}

Monodisperse, purely repulsive, hard spheres (MPRHS) provide an important model system for exploring fundamental mechanistic underpinnings of phase behavior in both atomic systems and colloids. Since the 1940s, phase transitions in these systems have been obtained or predicted via simulations, theory, and experiments. But there is an interesting gap in this literature: despite decades of exploration and reports of phase transition from one pure state to another, there have been no computational studies reporting spontaneous phase {\em separation} into coexisting domains of liquid and crystal regions. This gap owes its origin to the underlying mechanism of entropically-driven phase separation in MPRHS --- the competition between short-range (vibrational) entropy and long-range (configurational) entropy. Frenkel proposed that spontaneous phase separation in simulations of up to 1,000,000 particles would require more than 317,000,000 years to sample sufficiently many microstates to converge to a phase separated macrostate. Some brute-force simulations do show brief spontaneous coexistence but a metastable crystal or fluid subsequently overtakes the system. To bypass these difficulties, many studies use seeding, gravity, or direct construction of liquid and solid phases to study interfacial energy and nucleation rates of MPRHS systems. {WCA potentials have also been used to bypass metastability, where softness provides free volume, lowers osmotic pressure and the energy barrier}. It is well argued that the transition path taken in bypassing metastability is mechanistically the same as without triggers.  But as first acknowledged by Alder \& Wainwright, explicit observation of spontaneously-emergent coexistence is central to computational prediction of the first-order transition. Such observation would also provide satisfying demonstration of Frenkel's entropy exchange mechanism.   {After exploring the literature revealing these interesting behaviors, we conclude with an outlook for where to go next: computational demonstration of Frenkel’s mechanism for MPRHS awaits sufficiently large systems with carefully constructed hardness perturbations. }

\end{abstract}


\maketitle 

\section*{Introduction}
\label{sec:intro}
In colloidal dispersions, changes in packing fraction as well as changing strength of thermal energy, $kT$, relative to interparticle potential can induce transitions between liquid and solid phases, similar to phase behavior in molecular fluids. Here, $k$ is Boltzmann's constant and $T$ is the absolute temperature. How and under what conditions such phase transitions occur is very well studied, yet the fundamental model system underlying colloidal phase transitions --- purely repulsive, hard spheres --- produces surprising open questions, which we explore and address in this paper. In particular, we examine the extent to which the two conditions for a first-order phase transition have been observed in monodisperse, purely-repulsive hard spheres: phase \textit{transition} --- pure fluid and pure solid phases, and phase \textit{separation} --- explicitly coexisting fluid and solid domains \cite{alder1957phase, alder1959studies, alder1960studies}. Plots of osmotic pressure versus volume fraction or density underpin demonstration of this behavior.

We have reviewed the pioneering, landmark, and recent state-of-the-art literature demonstrating phase behavior in purely repulsive hard-sphere (PRHS) systems, in particular, for monodisperse particles. We take a closer look at a well-studied topic: colloidal phase separation driven by purely entropic forces, which is well-established for particles with shape anisotropy, and size-polydisperse spheres. However, we find that purely entropic phase separation for \textit{monodisperse}, purely repulsive hard spheres is not as completely established as we expected. {Specifically, many will find it surprising that no report of \textit{spontaneous}, \textit{equilibrium} fluid/crystal coexistence has been made in simulations.

As discussed below, phase transition and phase separation in monodisperse PRHS systems have been predicted via theory, simulations, and experiments dating back to the 1940s, where thermodynamic theory produces a tie line for the coexistence region, although only experiments have shown spontaneously-emerging explicit coexisting domains of fluid and crystal [\textbf{Figure \ref{fig:fig2}}]. Our inquiry was motivated by our own difficulty in simulating explicitly coexisting phases of fluid and crystal domains in colloidal suspensions of  up to 1,000,000 monodisperse, purely repulsive, very-hard spheres. In search of others' success with such systems, we were surprised to find no prior studies reporting equilibrium \textit{explicit}, coexisting fluid and crystalline phases in \textit{simulations}, except with the use of bias or strong triggers. {While several brute-force \cite{filion2010crystal, filion2011simulation} and event-driven molecular dynamics studies\cite{wohler2022hard} achieved nucleation, simulations are typically stopped at the nucleation point. The final state is unreported.  A few studies did continue after the nucleation event, invariably showing that the system is overtaken by a single phase.\cite{fiorucci2020effect, gispen2024finding} 

\begin{figure*}[t]
	\centering
	\includegraphics[width=0.9\linewidth]{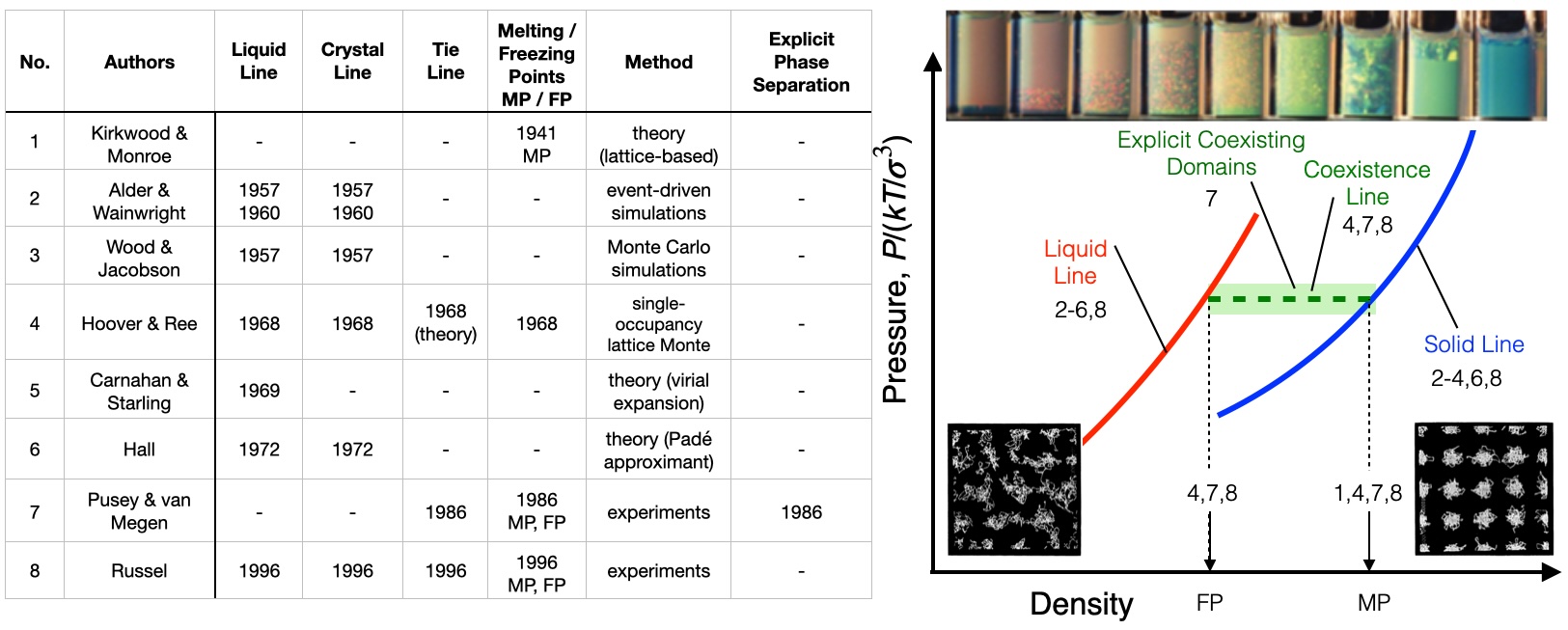}
	\caption{\textbf{Foundational atomic, colloidal, and rheological studies.} Phase transitions in systems of purely-repulsive hard spheres were established beginning with theory in the 1940s \cite{kirkwood1940theory,kirkwood1941statistical}, which matched experiments for argon \cite{eisenstein1940}. Alder and Wainwright's seminal event-driven simulations produced a fluid line and a crystal line \cite{alder1957phase,alder1960studies} as well as corresponding images (bottom insets) \cite{alder1959studies} and drove improvement of Wood and Jacobson’s Monte Carlo simulations \cite{wood1957preliminary}. Hoover and Ree combined lattice-constraint modeling with theory to obtain the phase lines and also used thermodynamic theory to deduce a tie line to put forth the hallmark melting and freezing points for monodisperse hard spheres \cite{hr-68}. Subsequent virial-expansion theory provided refinements \cite{cs-69,hall1972another,wertheim-63,mcquarrie-76}. Pusey and van Megen's landmark experiments with colloids showed explicit formation of coexisting fluid and crystal domains as shown in top inset \cite{pvM-86}. Russel and co-workers subsequently deduced colloidal pressure via x-ray measurements of packing, to produce phase lines and infer a coexistence line crossing gravitational layers \cite{prczcdo-96}.}
	\label{fig:fig2}
\end{figure*}

Aware that our system of 1,000,000 particles lacked competing entropy sources that easily trigger phase separation, such as shape anisotropy, size polydispersity, or softness, and that our model was built to match atomic theory, we wondered if other interparticle forces or increased system size would break the metastability.

The conventional energy/entropy competition mechanism we typically think about in colloids helps connect to atomic simulations and theory that had previously shown or predicted phase transitions. Experimentally-observed colloidal phases are sometimes mapped onto plots of interparticle attractions $kT/V_0$ versus volume fraction $\phi=4\pi a^3n/3$, producing a binodal as the gas/liquid phase envelope as well as melt and freeze lines (see, for example, a schematic illustration in Figure 1 in Padmanabhan \& Zia {\cite{pz-18}}) \cite{russel1991colloidal, lppsw-92, marr1993solid, evans1998metastable, sciortino2003evidence, zp-09}. Here, $V_0$ is the nominal interparticle interaction strength (and can be attractive, repulsive, or both), $a$ is particle size, and $n$ is the number density in solvent. Interactions range from strong adhesive to purely repulsive bonds. The phenomenology of these colloidal phase transitions is similar to that in molecular and atomic systems, where competition between attractions (which tend to condense) and thermal fluctuations (which tend to disperse) minimizes the system's total free energy \cite{mcquarrie-76, chandler1987introduction, balescu1975equilibrium}. More broadly speaking, these are different competing terms in the free energy. In molecular theory, for example, the Helmholtz free energy, $A$, encodes the competition between internal energy $U$ and entropy $S$ as $A=U-TS$ and can, for example, predict a liquid phase freezing into a crystalline solid, where the decrease in entropy is offset by a decrease of internal energy \cite{mcquarrie-76, chandler1987introduction, balescu1975equilibrium, frenkel-93}.  For both atomic and molecular systems, coexisting phases are thermodynamically connected to conditions of equal pressure and equal chemical potential via mean-field approaches such as (single-species) van der Waals theory \cite{van1873over} and (two-species) Flory-Huggins theory \cite{flory1942thermodynamics, huggins1942some}, which predict molecular phase transitions and coexistence, as well as regions of stability, instability, and metastability. Such theories require both internal energy and entropy for first-order phase transition and phase coexistence. 

First-principles simulations, free of assumptions of many theories, have also confirmed the energy/entropy competition mechanism for phase transitions. In a series of three seminal papers from 1957 to 1960, Alder and Wainwright tackled the problem of simulating many-body interactions in atomic systems, and demonstrated phase behavior with traditional energy and entropy competition. But their simulations also confirmed a then-controversial theoretical prediction from Kirkwood and Monroe \cite{kirkwood1941statistical}: phase transition driven by purely entropic forces with no obvious mechanistic competition \cite{alder1957phase,alder1959studies,alder1960studies}. 

Their event-driven simulations demonstrated phase transitions with monodisperse spheres, showing pure fluid and pure crystal phases, with either purely repulsive interactions or square-well attractions. Alongside this clear demonstration of phase {\em transition} in MPRHS, Alder and Wainwright highlighted that their small system could not produce explicit coexistence. Noting their desire to do so, they emphasized that it is worthwhile to seek explicit, spontaneous coexistence in dynamic simulations, stating: ``the larger systems were studied in the hope that the two states would eventually coexist. Only then can one be sure that hard spheres have a first-order phase transition between a fluid and a solid state.'' \cite{alder1960studies} We would add that not only is it important to observe such behavior, but also to report it.

Numerous subsequent studies have also produced fluid states and solid states \cite{hr-68, frenkel1984new, speedy1997pressure, wilding2000freezing, frenkel2002understanding, vega2007revisiting, odriozola2009replica, bannerman2010thermodynamic, nayhouse2011monte, fernandez2012equilibrium, statt2016monte, ustinov2017thermodynamics, pieprzyk2019thermodynamic, moir2021tethered}. As with molecular theories, these simulations-based, first-principles approaches have been successfully adapted to colloidal suspensions with {\em attractive interparticle forces} \cite{russel1991colloidal, lppsw-92, marr1993solid, evans1998metastable, sciortino2003evidence, zp-09}.  Subsequent colloidal simulations also demonstrate purely entropic phase transitions with shape anisotropy and size polydispersity, discussed below.

We traced this literature trajectory for development of atomic theories and simulations that specifically focus on PRHS, which set up the hallmark phase envelope for monodisperse hard spheres, with freezing and melting points $\phi_F=0.494$ and $\phi_M=0.545$ respectively, provided via intersection of the coexistence line of equal pressure and chemical potential with the phase lines. \textbf{Figure \ref{fig:fig2}} summarizes pioneering results, demonstrating how the atomic model has propagated into the colloids perspectives. All such approaches can predict the two pure phases that arise in first-order phase \textit{transitions} from one pure state to the other and, employing thermodynamic theory, deduce the coexistence line and region. The existence of the first-order phase transition in MPRHS colloids is well-established and not under debate. However, it is dissatisfying that no simulation studies have reported explicit, spontaneous \textit{separation} into fluid and crystal domains for monodisperse PRHS --- the same problem encountered by Alder \& Wainwright \cite{alder1957phase, wood1957preliminary, alder1959studies, alder1960studies}, as well as subsequent simulations of monodisperse, purely-repulsive hard spheres, including ours. Further, a clearer connection between the physics-based perspective and rheology perspectives would be beneficial to unifying understanding across these disciplines. We seek resolution of this dissatisfying situation, beginning with a closer look at models for purely-entropic phase separation.

\section*{Entropically-driven phase transitions and phase separation}
Onsager described purely entropically-driven phase transitions arising from shape anisotropy, which he demonstrated as a source of configurational entropy in the formation of liquid crystals of various highly anisometric particles \cite{onsager1949effects}. Much subsequent work built on these ideas for particle anisotropy \cite{eppenga1984monte, camp1997phase, cuetos2007kinetic, cinacchi2010phase, miller2010crystallization, kallus2011dense, agarwal2011mesophase, haji2011phase, jiao2011communication, avendano2012phase, marechal2012freezing, peroukidis2013phase, dijkstra2014entropy, boles2016self, karas2019phase, lim2023engineering}. Frenkel formalized the entropic \textit{competition} concept --- where shape anisotropy produces orientational entropy that can compete with translational entropy, which in turn drives \textit{e.g.}, isotropic-nematic phase transition in liquid crystals \cite{frenkel2001perspective}.

But even systems of spherical particles can undergo entropically-driven phase transitions, where the competition emerges from size polydispersity \cite{kranendonk1991computer, bartlett1992superlattice, eldridge1993entropy, han1994freezing, dijkstra1998phase, dijkstra1999direct, bw-99, fs-03, zubarev2005condensation, zaccarelli2009crystallization, wilding2010phase, hopkins2011phase, filion2011self, dijkstra2014entropy, boles2016self, koshoji2021diverse, koshoji2021densest}. Each size is a distinct species that introduces its own entropy term that competes with the others, producing fractionation and polycrystals \cite{mansoori1971equilibrium,Bartlett1990, bartlett-98, sollich1998projected,Phan1998, warren1998combinatorial,Kofke1999, bw-99,Bartlett2000, fs-03,Fasolo2004,pzvspc-09, zaccarelli2009crystallization, wilding2010phase, sollich2011polydispersity, bommineni2019complex, bommineni2020spontaneous}. Even the simple case of bidispersity leads to several phase behaviors, besides the entropic depletion attraction --- the Asakura-Oosawa potential \cite{asakura1954interaction, dijkstra1999direct, lekkerkerker2024phase}.

Confinement is also a source of size polydispersity, inducing heterogeneous nucleation near flat walls \cite{espinosa2019heterogeneous} and within a spherical cavity \cite{aponte2018equilibrium,wang2018magic, wang2019free, wang2020structural, sunol2023confined,mbah2023early}.  Mechanistically, the confinement length-scale is a second source of entropy that competes with particle size, inducing order at the wall to maximize entropy in the bulk \cite{aponte2018equilibrium, gonzalez2021impact, sunol2023confined}. Confinement is always a consideration for crystallization in experiments.

But, as touched upon in the Introduction, putatively \textit{monodisperse} purely repulsive hard spheres have long been observed to demonstrate phase transitions between a purely fluid and a purely crystal state, in both atomic and colloidal systems. However, monodisperse particles offer none of the previously-described shape-configuration or orientational entropy contributions, nor any attractive energy competition, indicating phase transition with no obvious \textit{competition} to drive the transition. While widely accepted as non-controversial today, it was controversial for at least two decades \cite{uhlenbeck1963p, ackerson1993order, frenkel-93}, because without an energy contribution, the crystal state would lead to a higher entropy than the fluid state.  Subsequent arguments, put forth by Frenkel in his 1993 letter \cite{frenkel-93}, suggested that configurational (long-range) entropy losses during freezing were offset by increased local vibrational (short-range) entropy, as atoms gain access to an entire local cage.
\vspace{-1mm}

\section*{Frenkel proposed a mechanism to explain monodisperse PRHS phase transition}

\vspace{-2mm}
Frenkel proposed a mechanism to explain the well-established predictions and observations of purely entropic phase transition in monodisperse, purely repulsive hard spheres.  Pointing to the apparent paradox of a higher-entropy crystal noted above, Frenkel highlighted Alder and Wainwright's simulations that first showed distinct phases: ``...in the 1950s, computer simulations indicated that a fluid of hard spheres could freeze. Hard spheres do not have any interactions so the potential energy of such a system is always zero.''\cite{frenkel-93}  Despite abundant reports of transition from one pure state to another \cite{kirkwood1940theory, kirkwood1941statistical,wood1957preliminary, alder1959studies, alder1960studies, hr-68, hall1972another, frenkel1984new, speedy1997pressure, speedy1998pressure, wilding2000freezing, frenkel2002understanding, kolafa2004accurate, vega2007revisiting, odriozola2009replica, bannerman2010thermodynamic, nayhouse2011monte, fernandez2012equilibrium, statt2016monte, ustinov2017thermodynamics, pieprzyk2019thermodynamic, moir2021tethered} and reports of experimentally observed coexistence \cite{pvM-86, prczcdo-96, rutgers1996measurement, zhu1997crystallization, hw-09}, the fact that there was no second species to provide entropy competition in \textit{monodisperse} hard-sphere systems left these phase transitions mechanistically unexplained.  Frenkel proposed that competition between vibrational and configurational entropy drives such phase transitions in monodisperse PRHS. {These concepts have been adapted in the colloids literature, where the former is described as short-range entropy associated with short-time self-diffusion within a nearest-neighbor cage, and the latter is termed long-range entropy, associated with long-time self-diffusion as a Brownian particle wanders through many suspension configurations.}

However, for \textit{monodisperse} hard-sphere systems, observation of phase \textit{transition} from a single-phase fluid to a single-phase crystal has proven much easier than direct observation of \textit{coexisting} phases --- phase \textit{separation} into coexisting domains of fluid and crystalline phases. In particular, simulations of monodisperse PRHS colloidal systems resulting in \textit{spontaneously}-formed, \textit{equilibrium} coexisting states of both phases have remained elusive, except with use of triggers that lower the metastable energy barrier, an informative topic discussed below. 


\section*{Spontaneous, equilibrium phase coexistence in MPRHS}		
Phase \textit{transition} and phase \textit{coexistence} in monodisperse, purely-repulsive hard-sphere systems are well-predicted by theory and confirmed in experiments that produce both the distinct, pure phases for transition and the explicitly coexisting phases for separation [\textbf{Figure \ref{fig:fig2}}]. In simulations of MPRHS, phase \textit{transition} is also confirmed via observation of distinct, pure phases. The \textit{coexistence} region in simulations is rigorously deduced as a tie line via thermodynamic theory. Many such simulations have been built to replicate the mono-disperse PRHS theory and putatively match corresponding experiments. So it surprised us that, while these simulations successfully predict distinct phases, none report explicit coexistence (separate phases).   The only exceptions reported require biased sampling of the free energy landscape in Monte Carlo simulations \cite{auer2001prediction, auer2004numerical, filion2010crystal, filion2011simulation, isobe2015hard}, or artificial construction of crystal substrates, e.g., for nucleation rate studies \cite{auer2001prediction, auer2004numerical, hermes2011nucleation, espinosa2014mold, espinosa2016seeding, tateno2019influence, fiorucci2020effect, gispen2024finding}; that is, they do not demonstrate \textit{spontaneous} emergence of a fluid/crystal coexisting state. While phase separation in MPRHS is not under debate, the condition sought is spontaneous (rather than triggered) emergence of equilibrium (durable) coexisting fluid and solid domains in simulations of monodisperse, very hard spheres. If this has been observed, so far, no one has reported it for any part of theoretical coexistence region. There seems to be a vague anecdotal consensus that such behavior has been observed, but report of observations is of course an indispensible part of the burden of proof for supporting a scientific claim. 

\section*{Single phases are predicted by theory and observed in simulations}
Liquid/solid transition in monodisperse spheres with general interparticle potential was predicted based on free-volume arguments as early as the 1930s by Lennard-Jones and Devonshire \cite{lennard1939criticala, lennard1939criticalb} and later improved in the early 1940s by Kirkwood and Monroe \cite{kirkwood1940theory, kirkwood1941statistical, kirkwood1950critique, kirkwood1951crystallization}. Kirkwood and Monroe's lattice-based thermodynamic model for the energy and entropy terms was derived using continuous distribution functions \cite{kirkwood1940theory, kirkwood1941statistical}, which expanded the prior lattice cell model \cite{lennard1939criticala, lennard1939criticalb} that always carried residual long-range order into the liquid state. By adopting the face-centered cubic (FCC) lattice, their model was reduced to a finite, tractable set of transcendental equations for the Fourier coefficients of the distribution functions, and the critical condition for a liquid-to-solid transition was predicted. They applied their reduced model to argon at $83.9K$ (near the experimentally measured melting temperature), a system typically treated as a PRHS fluid because the effect of attractive forces and energy is widely regarded as inconsequential in atomic fluids at high density and low temperature. With experimental data and approximation of interparticle potential as an 11.4/6 Lennard-Jones potential, they successfully predict the melting parameters (entropy of fusion, volume) for argon at several values of the temperature and pressure. Subsequent cell theories improved on the free volume calculation for strictly hard spheres \cite{buehler1951free, wood1952note} and for more general lattices \cite{squire1961formulation}. However, because these theories involve complicated integral equations that cover a massive configurational space, accurate computation of the equation of state was infeasible and could only be analytically solved on simple geometries such as an FCC lattice. As a result, these early free-volume theories obtained only the melting point. 

Computer simulations became feasible in the 1950s and moved the exploration of PRHS phase transition forward. Alder and Wainwright first observed distinct fluid and crystal phases in their pioneering simulations of monodisperse, purely repulsive hard spheres \cite{alder1957phase,alder1959studies,alder1960studies}, subverting previous models for disorder-to-order transitions where, as noted by Frenkel decades later \cite{frenkel-93}, ``it was commonly thought that attractive forces between molecules are essential for crystallisation: a crystal can form because the lowering of the potential energy of the system upon solidification `pays for' the decrease in entropy.'' While not the earliest work demonstrating entropically driven phase transitions in MPRHS, Alder and Wainwright's work has been perhaps one of the most impactful, cited by many others as foundational to their theory development. Their work produced fluid and solid lines \textit{and the open issue of no coexistence condition}. The authors attributed this situation to small system size (500 particles) \cite{alder1960studies} but whether the observation was a manifestation of proper metastability or just very small system size, was left open for future advances with sufficient volume to properly test for metastability.

Later expansions of the theory bypassed the competing-forces mechanism, instead using phenomenological (virial) expansions of equations of state relating pressure and density, rather than free-energy minimization. Many such approaches predict the equilibrium fluid line \cite{thiele1963equation, wertheim1963exact, mcquarrie-76, ree1964fifth, ree1967seventh, cs-69, clisby2006ninth, schultz2014fifth} and, separately, others predict an equilibrium solid line \cite{hall1972another}. These fluid and solid lines are illustrated in \textbf{Figure \ref{fig:fig2}}, as well as a dashed line indicating coexistence regions deduced by theory (that uses parameters obtained from simulations). While many approaches calculate a fluid line, relatively few theoretical approaches have predicted a solid line in atomic systems. Hall obtains a solid line using an \textit{ad hoc} revision of fluid state theory for PRHS \cite{hall1972another}, as well as a fluid line.  

Combining this new theory with advanced computational approaches finally led to the complete phase envelope. Hoover and Ree built PRHS atomic theory into simulations, using virial expansions to obtain the fluid line. For the solid region, they used single-occupancy lattice modeling and Monte-Carlo simulations to calculate the pressure. They then integrate the pressure along a reversible path, starting with a reference state with known Helmholtz free energy (ideal gas or Einstein crystal), to obtain the free energy at the target state. The entropy and chemical potential are then derived from the free energy for an individual phase. Competing forces may be present {\em within a single phase}, but they do not constrain one another to produce a second phase or to directly identify the coexistence line. 

For the solid-line construction, the method does impose a free energy contribution to constrain the model's lattice (the hard spheres have no interparticle attractions). By design, this constraint prevents emergence of coexisting fluid. In light of Frenkel's subsequent assertion, we view the lattice constraint as a configurational constraint, competing with the entropy of particles' long-range entropy. This constraint appears as a term in Hoover \& Ree's free energy expression, providing a source of competing potentials in their model. Thus, they generate a fluid line and a pure crystal phase, but no phase separation into coexisting domains (by construction of the lattice constraint). To deduce the coexistence region, the authors calculate points of equal chemical potential in the fluid and crystal states and draw the tie line between the two points. This rigorous result was one of the earliest to definitively identify the coexistence region. With this approach, they identify the freezing and melting transitions at 49.4\% and 54.5\% volume fraction, respectively. This landmark result is used nearly universally in the literature as the phase envelope for monodisperse purely repulsive hard spheres. 

Many subsequent advancements in computational methods have been used to refine the phase envelopes for MPRHS, including Monte Carlo methods \cite{frenkel1984new, wilding2000freezing, frenkel2002understanding, vega2007revisiting, odriozola2009replica, bannerman2010thermodynamic, nayhouse2011monte, fernandez2012equilibrium, statt2016monte, ustinov2017thermodynamics, moir2021tethered}, molecular dynamics simulations \cite{speedy1997pressure, speedy1998pressure,kolafa2004accurate, pieprzyk2019thermodynamic}, Brownian dynamics \cite{fb-00}, and Stokesian dynamics \cite{swaroop-10}. While this abundance of simulation studies reinforces the first-order transition and phase envelope for MPRHS, none report spontaneous, explicit fluid-and-crystal coexistence. There are many studies that \textit{induce} explicit phase coexistence via seeding and other methods, or which produce short-lived coexistence that gives way to a metastable fluid or crystal. These are discussed in detail below in the section \textit{Nucleation and phase separation without and with triggers}. But experimental studies have reported direct observation of \textit{coexisting states}, which we discuss next. 

\section*{Experiments report phase transition and explicit phase separation}
\vspace{-3mm}
Experimental systems approach to greater or lesser extent the pristine MPRHS condition. Nearly hard-sphere, nearly monodisperse, purely-repulsive colloidal systems report phase transitions and explicitly observe fluid/crystal coexistence \cite{pvM-86, pvM-87}. Plots of crystal fraction versus volume fraction help establish phase boundaries as extrapolations of coexistence data. Pusey \& van Megen's seminal study mapped out the distinct phases and phase coexistence, providing estimates of the melting and freezing points by extrapolating the coexistence points \cite{pvM-86}. The core particle size from dry measurements inevitably under-predicts the atomic theory melting point, due to particle swelling. To compensate, they manually adjusted their freezing point to agree with Hoover and Ree's atomic theory (49.4\%), pointing also to a combination of particle softness, interparticle attractions, and size polydispersity. The rescaling also moved their melting point to volume fraction 53.6\%. This, and other, necessary adjustments became standard practice in experiments and colloidal simulations using a soft interparticle potential, including careful particle-size adjustments \cite{rpw-13}. As a result of these unavoidable particle conditions, freezing and melting points for colloids requires care to match to atomic theory \cite{poon2012measuring}. But overall, experiments demonstrate phase coexistence, where particle softness and size polydispersity shift the phase envelope. The colloids and rheology community thus smoothly adapted the phase envelope developed by atomic theory and, in fact, provided much of the prominent experimental demonstration, both in terms of structure and osmotic pressure.

This impact of soft potentials on experimentally-obtained phase envelopes has been characterized extensively in simulations. Many colloidal simulations employ soft potentials such as the Lennard-Jones or Weeks-Chandler-Anderson (WCA) potential and rescale their data to match the freezing point \cite{filion2011simulation, tateno2019influence, fiorucci2020effect}. A robust literature reports how this softness, as well as elasticity, affect colloidal packing and phase behavior \cite{hoover1971thermodynamic, Robbins1988, Meijer1991, piazza1993equilibrium, Lowen1993, Lowen1993b, nemeth1995solid, senff1999temperature, likos2001effective, castelletto2002liquid, Laurati2005, archer2005density, likos2006soft, mladek2007phase, mladek2007clustering, Vlassopoulos2014, gupta2015dynamic, pelaez2015impact, Zakhari2017, erigi2023phase}, describing the shift of the freezing, melting, and maximum packing points. 

But the coexistence observed in experiments may not be strictly spontaneous, instead triggered by gravity, electrostatic effects, and container wall effects \cite{pvM-86}. Spontaneous appearance of an ordered phase (from the metastable fluid state) requires a nucleation event, and these are rare, so it can take a long time for the system to sample strong enough fluctuations to get pushed out of metastability. In experiments, gravity and crystal seeding (via sample tumbling) are thought to artificially trigger the system out of metastability \cite{vMu-94, prczcdo-96, rutgers1996measurement, hermes2011nucleation}. Russel and co-workers highlighted these phenomena explicitly \cite{prczcdo-96, rutgers1996measurement}.  In their experiments, gravity settled a thick crystal layer at the bottom with a colloidal fluid phase above it; the interface was loosely identified as a coexistence region. The authors then used this to sketch a tie line into the resulting pressure plot (see Phan \textit{et al.}'s Figure 5 \cite{prczcdo-96}). Careful confocal microscopy experiments by Weeks and co-workers quantified many features within such a liquid/crystal interfacial layer in their own experiments (see \textbf{Figure \ref{fig:fig3}} above), 
\begin{figure}[t]
	\centering
	\includegraphics[width=0.9\linewidth]{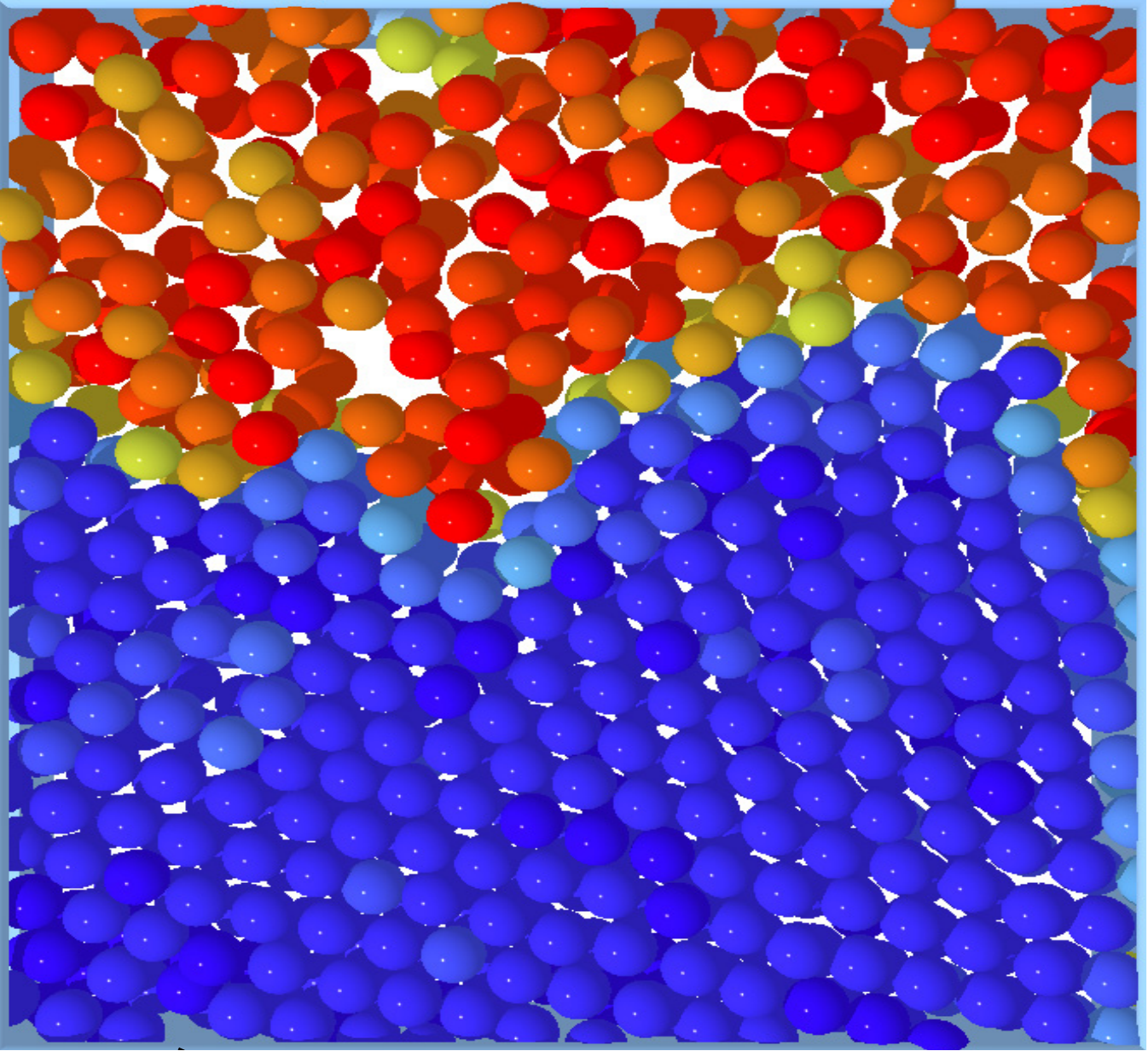}
	\caption{Confocal microscopy images from \cite{hw-09} showing the interface between colloidal liquid and crystal regions in a suspension of PMMA particles in cyclohexylbromide and decalin. Particles colors indicated structure, from dark blue for crystals to red for liquid. With permission from \textit{Proc. Nat. Acad. Sci.} }
	\label{fig:fig3}
\end{figure}
noting that particle density mismatch assisted this gravitational separation \cite{hw-09}. The authors also point out that weak electrostatic interactions contribute in the form of particle size, which drives rescaling. A later confocal microscopy study of charged colloids showed a narrow range of fluid-BCC coexistence and identified large-scale, correlated thermal fluctuations to stabilize BCC crystal near melting \cite{sprakel2017direct}, but electrostatics provide an enthalpic contribution to Helmholtz free energy. Between these studies, Russel \& Chaikin \textit{et al.} \cite{zhu1997crystallization} studied gravity's effects on colloidal crystal formation by monitoring changes in a suspension of PMMA beads with 10nm soft repulsion and 5\% size polydispersity, using a sample prepared on Earth and then evaluated on the Space Shuttle Columbia. The study focused on crystal structure and formation, rather than crystal/liquid phase separation. A sample prepared on Earth at $\phi= 0.505$ produced a phase-separated suspension of 0.1mm crystallites gravity-settled to the bottom with a supernatant layer above, recovering the observations made by Pusey \& van Megen in a very similar system \cite{pvM-86}. The sample was then shear-rejuvenated on the Space Shuttle. Whether the shear rejuvenation protocol  restored the system to a metastable liquid was not reported. After three days in flight, the crystallites, now distributed homogeneously throughout the liquid phase, had grown to millimeter-sized crystals with dendritic arms. The authors reported that gravity contributes to the mixture of RHCP and FCC crystal structure, along with dendritic arms arising from an instability also observed in molecular systems \cite{langer1980instabilities, russel1997dendritic}. The authors also remarked that the competition between sedimentation flow and Brownian motion during settling affects formation of dendritic arms and crystal structure. A subsequent study of 7\% size polydisperse, very hard particles showed explicit coexistence using scanning electron microscopy (SEM), but again crystallization was aided by gravity pushing particles into an ordered configuration on a flat surface. \mbox{\cite{jiang1999single}} Overall, obtaining phase separation without gravitational settling is satisfying and eliminates one factor from experiments that obfuscates Frenkel's proposed mechanism, and also suggests the use of neutrally buoyant particles in colloidal simulations. 

Altogether, experiments do achieve fluid and crystal phase separation and coexistence. The presence of size polydispersity and seeding play a role beyond the entropic forces that drive phase separation in an ideal monodisperse PRHS suspension. This provides a useful comparison to the pristine conditions in simulations that fail to exit metastability, discussed below.

\section*{Nucleation and phase separation, without and with strong triggers}
\begin{figure*}[t]
	\vspace{0mm}
	\centering
	\includegraphics[width=0.9\linewidth]{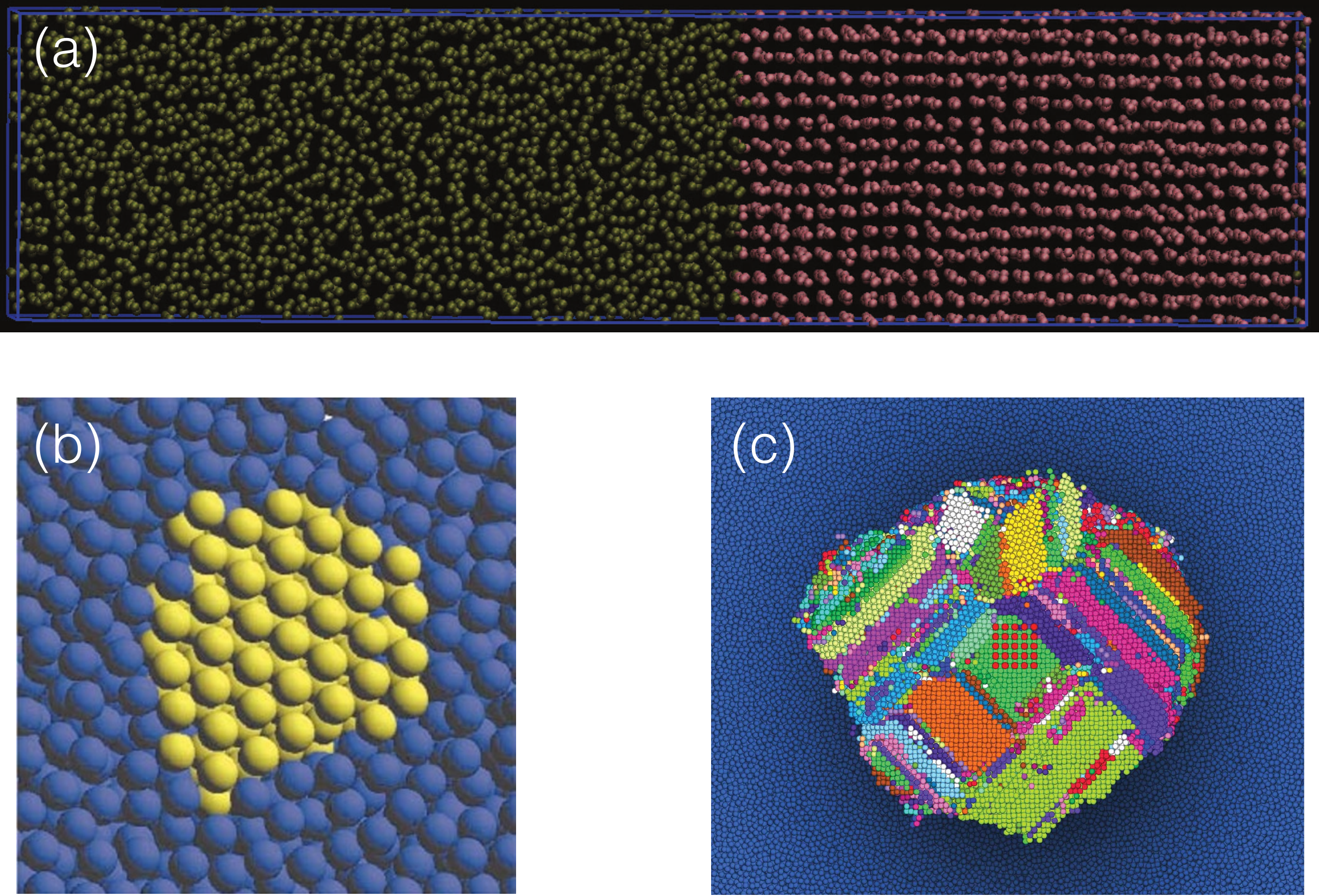}
	\caption{Methods for triggering crystal nucleation in simulations. (a) The ``direct coexistence" method constructs an equilibrated face-centered cubic phase (right half of simulation cell, 2,548 particles) and a separate equilibrated fluid phase (left half of cell, 2,548 particles), pressing them into contact as the initial configuration. Figure from \cite{espinosa2013fluid}, with permission from \textit{AIP/ Journal of Chemical Physics}. (b)  Biased-sampling Monte Carlo method: snapshot of transient critical nucleus at $\phi=0.5207$ (yellow), surrounded by fluid-phase particles (blue). The nucleus is pre-constructed using biased Monte Carlo simulations to speed the nucleation process. Figure from \cite{auer2001prediction}, with permission from \textit{Nature}. (c) Seeding method: snapshot of molecular dynamics simulation illustrating a $5 \times 5$ square seed promoting surrounding crystal growth. From \cite{hermes2011nucleation}, with permission from Royal Society of Chemistry, \textit{Soft Matter}.}
	\label{fig:fig4}
\end{figure*}

Nucleation is notoriously slow in simulations, making spontaneous coexistence quite difficult to achieve without the massive number of microstates available in experiments. Waiting long enough for the metastable state to sample strong enough fluctuations to push the system out of metastability is thus a key challenge. {According to classical nucleation theory (CNT), once a nucleus larger than the critical size is formed, the crystal growth progresses indefinitely, so }the most probable final state following nucleation from a metastable fluid state, without other triggers, is a full crystal. With `small' simulation size, this is the overwhelmingly most probable final macrostate. Dijkstra and co-workers thoroughly studied nucleation rates without strong triggers for a system of atomic hard spheres using event-driven simulations \cite{filion2010crystal}. However, the simulations were stopped upon the first nucleation event, and the final state was not reported. Given an additional large number of events, it is probable the system would reach the metastable crystal state. In their subsequent Brownian dynamics simulations, colloids interacted via a WCA potential \cite{filion2011simulation}. The WCA potential results in a reduced virial coefficient value of $B_2^*=0.729$, which is quite soft, allowing significant volume adjustments. Soft potentials are well-known to lower the nucleation energy barrier \cite{hoover1971thermodynamic, senff1999temperature, likos2001effective, castelletto2002liquid, archer2005density, mladek2007phase, mladek2007clustering, pelaez2015impact}. In this case, the simulations were again stopped at the first nucleation event. In later work, Dijkstra and co-workers utilized brute-force dynamic simulations with the WCA potential, starting with a fluid structure in the coexistence region, then evolving dynamics up to 4,000 Brownian times over a range of volume fractions \cite{fiorucci2020effect}. Several runs were tested at each volume fraction (see Fiorucci, Dijkstra \textit{et al.}'s Figure 1a \cite{fiorucci2020effect}). For all volume fractions tested, most of them ultimately fully crystallized into the metastable state. The remaining runs showed continued growth of volume fraction over time, indicating that the metastable crystal state would again prevail. The appearance of temporary phase-separated domains is encouraging, although the use of the soft potential substantially lowers the metastable barrier. Most recently, they allowed an FCC crystal prepared in the coexistence region to melt to study nucleation of fluid packets \cite{gispen2024finding}. They noted that ``formation of a nucleus can take a long time, but once it reaches a critical size, it rapidly expands, leading to melting of the entire system'', i.e., a metastable fluid. W\"{o}hler and Schilling tested a much larger system with 1,000,000 particles \cite{wohler2022hard}. Their event-driven molecular dynamics simulations focused on nucleation with similar outcomes showing snapshots during full crystal growth; {The ultimate state of the system was not reported.} Notably, they resolved the discrepancy between experimental and simulation nucleation rates. We remark that applying the lever rule within any one of these {not-equilibrated} studies would result in widely different answers, including 100$\%$ pure solid or pure fluid, even though these studies examined systems within the coexistence region. An appropriate application of the lever rule requires an equilibrium (intransient) state, e.g.,  Pusey and van Megen's Figure 2 in their 1986 study \cite{pvM-86}. Waiting a long time in the above simulations would move the data to the pure phase rather than produce a lever-rule line. Overall, there have been no reports of {\em spontaneous} coexistence states for MPRHS.

The need to trigger phase separation reflects the metastability of MPRHS in the region $0.494 \le \phi \le 0.545$. Even simulations reporting of {\em spontaneous nucleation} in hard sphere systems is restricted to a narrow range: Filion {\em et al.} and W\"{o}hler {\em et al.} reported spontaneous nucleation only for $\phi > 0.53$ \cite{filion2010crystal, wohler2022hard}. \\

\noindent\textit{Direct coexistence}. Some approaches bypass the nucleation process altogether. ``Direct-coexistence'' methods \cite{ladd1977triple} pre-construct the crystal and liquid regions separately, then manually push them together (see \textbf{Figure \ref{fig:fig4}(a)}), typically to study the interface. Such systems are prepared close to the theoretical coexistence line and are allowed to equilibrate \cite{davidchack1998simulation, noya2008determination, zykova2010monte, espinosa2013fluid, tateno2019influence, sanchez2021fcc}. This manually-introduced interface has been obtained in atomic systems via Monte Carlo \cite{noya2008determination, zykova2010monte, bultmann2020computation} and molecular dynamics simulations \cite{davidchack1998simulation, espinosa2013fluid, sanchez2021fcc}, as well as in mesoscale colloidal \cite{tateno2019influence} simulations. For example, Tanaka's method pre-constructs a 3D crystal, inserts it in a dense colloidal fluid near the theoretical coexistence line, and allows Brownian hydrodynamics to equilibrate the system. 
	
	Many direct-coexistence simulations approximate hard spheres by using a relatively soft potential such as the WCA, and use particle-size or other rescaling to match theoretical phase lines. Examples include Tanaka's rescaling to match the atomic hard-sphere theory value of $\phi=0.494$ \cite{tateno2019influence}, then applying the adjustment to the entire data set. If one adjusts the particle size from that study to correspond to particles' thermodynamic radius, the model predicts a freezing point of $\phi_F\approx0.528$, consistent with the well-known shift of soft particles' phase envelopes \cite{hoover1971thermodynamic, piazza1993equilibrium, nemeth1995solid, senff1999temperature, likos2001effective, castelletto2002liquid, archer2005density, mladek2007phase, mladek2007clustering, pelaez2015impact}. 
	
	Overall, these approaches avoid metastability entirely. This avoids the real problem of dynamic accessibility and is expedient. However, direct-coexistence does not produce \textit{spontaneous} phase separation driven by competing forces into coexisting phases, the sought-after demonstration for MPRHS. 
	
	Alternatively, one can strongly perturb the system to lower the metastable energy barrier, and induce phase separation. \\

\noindent\textit{Crystal seeding method}. The nucleation rate and the interfacial free energy are attributes of the critical nucleus that will eventually grow into a final equilibrium state, including phase-separated states in the coexistence region. Many studies interested in nucleation rates aim to reduce the long wait by seeding a small substrate --- from a few to a few hundred particles --- upon which nucleation occurs. These nucleites are crystal seeds in either two or three dimensions. In all such approaches, the pre-constructed, pre-stabilized crystal cluster is co-located within a surrounding fluid phase, and nucleation is monitored. Monte Carlo simulations are one approach that successfully uses seeding to study nucleation rates, with embedded techniques such as umbrella sampling \cite{auer2001prediction, auer2001suppression, auer2004numerical, filion2010crystal, filion2011simulation, gispen2024finding}, forward flux sampling \cite{filion2010crystal, filion2011simulation}, and the mold integration method \cite{espinosa2014mold}. Frenkel and coworkers'  methods \cite{cichocki1990dynamic} produced measurements of crystal nucleus size and nucleation barrier distribution [\textbf{Figure \ref{fig:fig4}(b)}] \cite{auer2001prediction, auer2001suppression, auer2004numerical}.  But even for these approaches, no phase separation into an equilibrium coexistences state was reported, at any volume fraction.  

Alternatively, molecular dynamics simulations are used to better mimic particle dynamics. Seeded, unbiased molecular dynamics simulations also produce nucleation \cite{hermes2011nucleation, espinosa2016seeding, fiorucci2020effect, gispen2024finding}. For example, Dijkstra and co-workers manually inserted a fixed, ordered 2D array of particles [\textbf{Figure \ref{fig:fig4}(c)}] upon which a crystal grew, in both atomic simulations and colloidal experiments\cite{hermes2011nucleation}. {This produced a transient snapshot of a growing nucleus, but stopped short of establishing the equilibrium state built on the seeding platform}. Unfortunately however, these studies are typically ended as soon as the nucleation event is observed. One 2D seeding approach has achieved and reported triggered coexistence: the mold integration method \cite{espinosa2014mold} starts with a pre-constructed 2D crystal slab that exerts square-well attractions on nearby particles arranged in a lattice plane, seeding the accumulation of more crystal structure around the slab. {The induced crystalline structure bypasses the metastable state, facilitating calculation of interfacial free energy between the coexisting liquid and crystal domains}. But the path followed by the triggered nucleation event is believed to be the same as that would be followed by the spontaneous event, as claimed by Espinosa \textit{et al.} \cite{espinosa2016seeding} who, among others, used 3D crystal seeds to trigger nucleation \cite{espinosa2016seeding, fiorucci2020effect, montero2020young, montero2020interfacial, gispen2024finding, Richard2018a, Richard2018b}.  \\

Altogether, these simulations either bypassed the metastable state entirely (direct coexistence) or partly. For the latter, the two system conditions used to reliably bypass metastability are particle softness and triggers such as gravity and seeding. But the ultimate outcome of the above studies is either left unknown because the simulation is stopped upon nucleation \cite{auer2001prediction, auer2001suppression, auer2004numerical, filion2010crystal, filion2011simulation, hermes2011nucleation, espinosa2016seeding, montero2020young, montero2020interfacial, Richard2018a, Richard2018b}, or the ultimate outcome is that the system is overtaken by a single-phase metastable state \cite{fiorucci2020effect, gispen2024finding}. That is, they produce no reports of \textit{spontaneous} emergence of phase separation. Together these reinforce the fact that the MPRHS coexistence region is metastable. \\

But somehow, phase separation occurs in experiments in finite time. The large number of particles in an experiment evidently facilitates finite-time sampling of the phase-separated macrostate. One can use experimental data to estimate how long this would take in simulations, as a function of simulation system size. It has been predicted that even with 1,000,000 particles it would take 317,000,000 years, about 2$\%$ of the life of the universe, to generate spontaneous, durable phase separation in simulations of MPRHS \cite{tenwolde1996numerical}. But colloidal phase separation in experiments occurs without the pristine conditions of a strictly monodisperse, neutrally-buoyant hard-sphere system, suggesting that simulation time could be even longer. That is, it is actually unsurprising that such phase separation has not --- and cannot have --- been reported for truly MPRHS systems. This difference in conditions also suggests which pristine conditions to minimally perturb to sample phase separation in finite time.

\begin{figure}[t]
	\centering
	\includegraphics[width=0.9\linewidth]{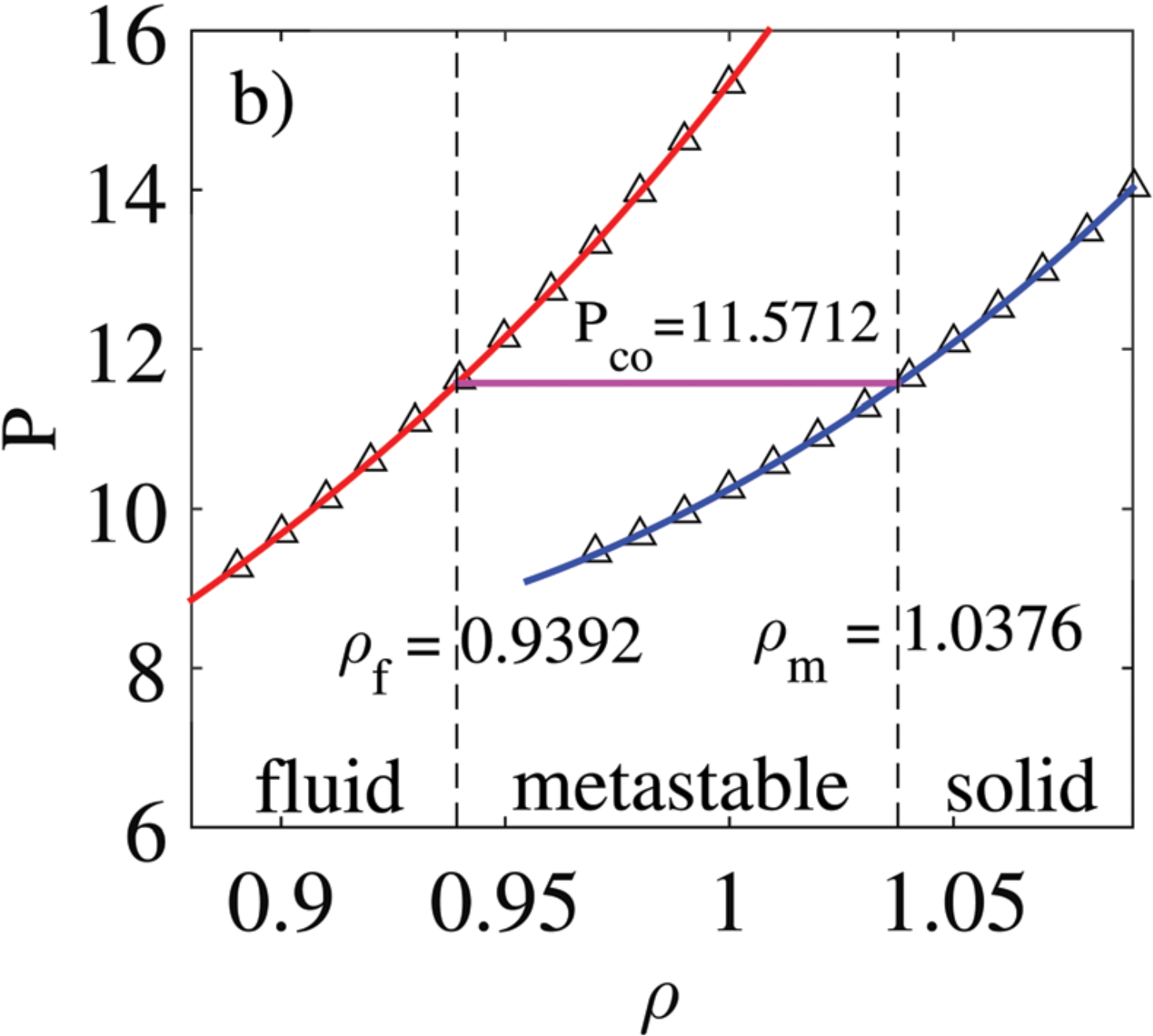}
	\caption{Pressure $P$ as a function of number density $\rho$ spanning fluid, coexistence and solid regions from \cite{pieprzyk2019thermodynamic}. Pressure scaled with $kT/\sigma^3$, where $kT$ is the thermal energy and $\sigma$ is particle diameter. Number density is related to volume fraction as $\phi =  \rho \pi \sigma^3/6$. Coexistence tie line obtained by equal-pressure and equal-chemical-potential conditions, giving $\phi_F=0.492$ and $\phi_M=0.543$. With permission from the Royal Society of Chemistry, \textit{Physical Chemistry Chemical Physics}.}
	\label{fig:fig5}
		\vspace{-6mm}
\end{figure}

\section*{Recent advances: large-scale, long-duration simulations with truly hard spheres}
\vspace{-3mm}
The metastability that naturally prevents spontaneous fluid-and-crystal coexistence from emerging as a distinct, explicit phase in simulations does not call into question whether the first-order transition exists. The entropically-driven fluid/solid phase transition has been clearly predicted in theory and obtained in experiments and simulations. Frenkel's subsequent mechanistic explanation of the driving force in MPRHS also automatically predicts that we will never see phase coexistence in finite time in pristine simulations built to replicate atomic theory. At the heart of his explanation and this difficulty is the Law of Large Numbers. Simulations are just too small to sample sufficiently many microstates to converge to a phase-separated macrostate in humanly measurable time. But we are still interested in finding the least perturbation needed in a very large simulation.  Indeed, the stated aim of Alder and Wainwright's 1960 study was to establish the minimum system size required to achieve phase coexistence, but the limited compute power of that time capped their study at 500 particles \cite{alder1960studies}. The authors ultimately found it unsurprising that 500 particles was woefully undersized, and anticipated future computational advances to simulate large enough systems to spontaneously sample a coexistence region for a pristine system.

To wit, even Pieprzyk \textit{et al.}'s recent large-scale, long-duration study of truly hard spheres {did not report} explicit phase separation. Using event-driven molecular dynamics with 1,000,000 truly hard particles, they successfully produced pure fluid and pure crystal states [\textbf{Figure \ref{fig:fig5}}]. They used these phase lines and careful calculations of chemical potential to theoretically deduce a coexistence line \cite{pieprzyk2019thermodynamic}. The authors developed rigorous theory (expanding the work of Kolafa \textit{et al.} \cite{kolafa2004accurate} and Speedy \cite{speedy1998pressure}) to calculate chemical potential and recover (and refine) the Hoover and Ree hallmark volume fractions for hard-sphere melt and freeze points. A tie line between them indicates a coexistence region, but simulations in that region revealed only purely fluid or purely crystal structure --- no {report of} coexisting domains. The authors attribute this to metastability. We agree. We speculate that it is also related to their use of infinitely hard particles undergoing ballistic collisions.

\vspace{-3mm}
\section*{Outlook}
\vspace{-3mm}

 \begin{figure*}[t]
	\vspace{-0mm}
	\centering
	\includegraphics[width=0.95\linewidth]{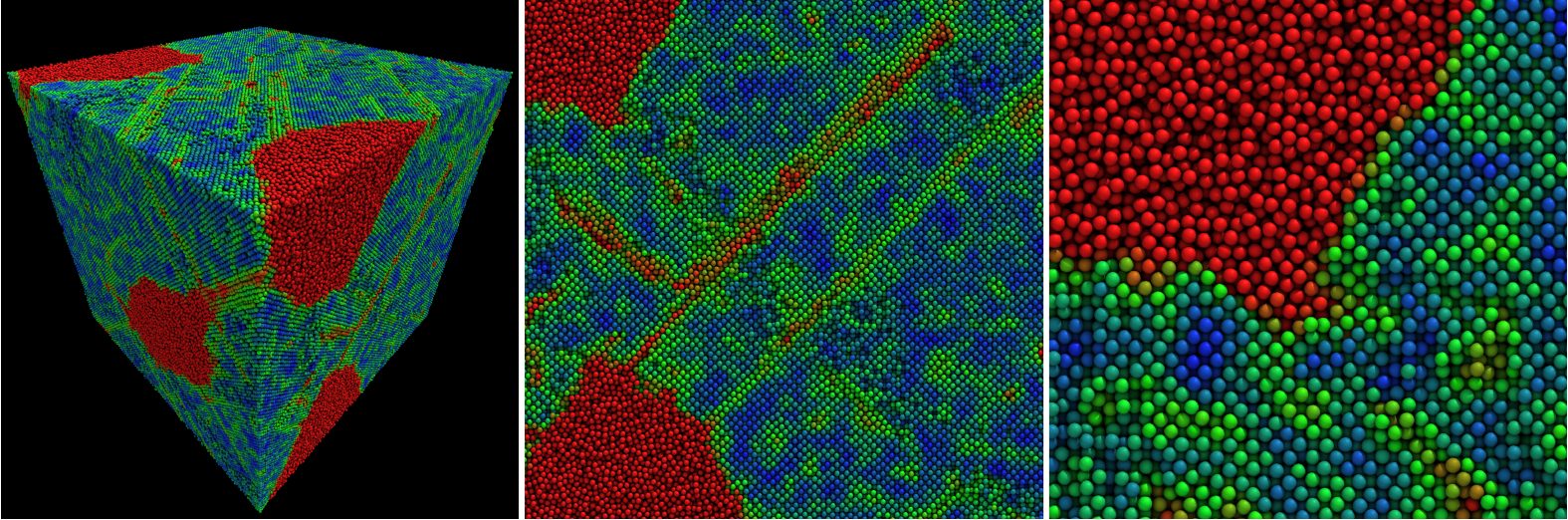}
	\caption{Recent large-scale Brownian dynamics simulations from \cite{wangInprepDemonstration}. Far left: simulation cell of 2,000,000 colloids, replicated periodically into an infinite domain in LAMMPS \cite{thompson2022lammps}. Second and third images: same system at 2x and 5x magnification. Colors correspond to local order, ranging from red for structureless to deep blue for perfect crystal structure. With permission, \textit{J. Chem. Phys.}}\label{fig:fig6}
	\vspace{-6mm}
\end{figure*}
We have reviewed the pioneering, landmark, and recent state-of-the-art literature demonstrating phase behavior in monodisperse, purely repulsive hard-sphere systems. Our inquiry was motivated by our own difficulty in triggering explicit phase separation in large-scale colloidal suspensions of purely repulsive, monodisperse, very-hard spheres. In search of other simulation studies of the same type of system, we were surprised to find no prior such studies where equilibrium explicit, coexisting fluid and crystalline phases were reported.  Recognizing that our system of 1,000,000 particles lacked shape anisotropy, size polydispersity, or softness --- familiar sources of competing entropy that readily trigger phase separation --- and that our model was built to match atomic theory, we wondered if other interparticle forces or increased system size would break the metastability. 

Here we traced a trajectory in the literature following the development of atomic theories and simulations that specifically focus on PRHS, which set up the hallmark phase envelope for hard spheres, with freezing and melting points $\phi_F=0.494$ and $\phi_M=0.545$ respectively, provided via intersection of the coexistence line of equal chemical potential with the phase lines, which follows the equilibrium thermodynamics. Decades of subsequent simulations make it clear that this region is metastable, mirrored in the many studies of nucleation rate. 

All simulations of monodisperse PRHS reviewed can predict first-order phase transitions from one pure state to the other, but none reported explicit, equilibrium, spontaneous formation of fluid and crystal domains  --- the same problem we had encountered. {Previous studies claiming to observe fluid/crystal coexistence did not meet one or more characteristics of the pristine conditions of the atomic model: particles were not spherical (shape anisotropy) \cite{onsager1949effects, eppenga1984monte, camp1997phase, cuetos2007kinetic, cinacchi2010phase, miller2010crystallization, kallus2011dense, agarwal2011mesophase, haji2011phase, jiao2011communication, avendano2012phase, marechal2012freezing, peroukidis2013phase, dijkstra2014entropy, boles2016self, karas2019phase, lim2023engineering}, not monodisperse (polydispersity) \cite{kranendonk1991computer, bartlett1992superlattice, eldridge1993entropy, han1994freezing, dijkstra1998phase, dijkstra1999direct, bw-99, fs-03, zubarev2005condensation, zaccarelli2009crystallization, wilding2010phase, hopkins2011phase, filion2011self, dijkstra2014entropy, boles2016self, koshoji2021diverse, koshoji2021densest, bommineni2019complex, bommineni2020spontaneous}, or not hard (e.g., $0.7 \leq B_2^* \leq 0.8$) \cite{filion2011simulation, fiorucci2020effect, gispen2024finding, espinosa2013fluid, espinosa2016seeding, montero2020young, montero2020interfacial, sanchez2021fcc, boles2016self, espinosa2019heterogeneous, bultmann2020computation}; or, the phase separation was not spontaneous (biased Monte Carlo, direct coexistence, crystal seeding) \cite{vega2007revisiting, noya2008determination, filion2010crystal, filion2011simulation, fiorucci2020effect, espinosa2013fluid, espinosa2014mold, espinosa2016seeding, montero2020young, montero2020interfacial, sanchez2021fcc, boles2016self, wang2018magic, wang2019free, wang2020structural, mbah2023early, espinosa2019heterogeneous, isobe2015hard, bultmann2020computation} and, typically, the system was not equilibrated  \cite{filion2010crystal, filion2011simulation, fiorucci2020effect, wohler2022hard, gispen2024finding}. }

While experiments have long demonstrated phase transition and phase separation, authors also remark that gravity, size polydispersity, and other particle perturbations break  metastability to yield phase separated samples, largely confirming the perturbations used in simulations. 

But these results are not surprising.  Breaking MPRHS metastability in pristine simulations that replicate atomic theory is overwhelmingly improbable with finite system size. Frenkel's mechanistic model underscores the difficulty of satisfying the Law of Large Numbers: the competition between long-range (configurational) and short-range (vibrational) entropy underlying phase transitions in monodisperse PRHS systems requires a huge system or very, very long times to sample sufficiently many microstates to converge to a phase-separated macrostate.

Setting aside strong triggers that essentially bypass the metastable energy landscape, it would be worthwhile to identify the minimal perturbation needed in a tractably large simulation to explicitly traverse the coexistence domain in a reasonable time frame.

Future simulation studies should obviously be larger than prior studies. The ability of particle softness to easily trigger phase separation suggests that future investigations formulate a minimal perturbation to particle interactions, which has been shown to strongly affect nucleation rates. We have begun to test this idea and find promising preliminary results, as shown in \textbf{Figure \ref{fig:fig6}} \cite{wangInprepDemonstration}, by systematically increasing system size and testing values of the reduced second virial coefficient closer to unity. Further connections between the physics and rheology perspectives will be useful as well, including detailed studies of how osmotic pressure changes with particle conditions, thus affecting phase behavior.

\section*{Acknowledgments}
The authors acknowledge the support of the National Science Foundation's computational resources: Anvil at the Purdue University's Rosen Center for Advanced Computing (RCAC) \cite{Anvil} and Ranch Storage at Texas Advanced Computing Center (TACC) at U.T. Austin  through allocation CHM240060 from the \mbox{ACCESS} program \cite{ACCESS}, which is supported by U.S. National Science Foundation grants \#2138259, \#2138286, \#2138307, \#2137603, and \#2138296. JGW acknowledges helpful conversations with Dr. Gesse Roure. The authors gratefully acknowledge insightful feedback from anonymous reviewers, if not the pain involved, which led to substantial improvement and expansion of the reported work.

\section*{Declaration of Competing Interests}
The authors declare that they have no known competing financial interests or personal relationships that could have appeared to influence the work reported in this paper.

\section*{Data Availability Statement}
Data are stored on the Ranch Storage at Texas Advanced Computing Center (TACC) at U.T. Austin and are available upon request.

\clearpage

\end{document}